\documentclass[conference,10pt]{IEEEtran}
\usepackage{epsfig,epsf,rotating,setspace,latexsym,amsmath,amssymb,amsfonts,bm,theorem,subfigure,epstopdf}
\usepackage{cite,authblk}
\usepackage{bbm}
\usepackage{algorithm}
\usepackage[noend]{algpseudocode}
\usepackage{color}
\usepackage{mathtools}
\usepackage{soul}

\algrenewcommand\algorithmicforall{\textbf{foreach}}
\algrenewcommand\algorithmicindent{.8em}

\IEEEoverridecommandlockouts
\allowdisplaybreaks

\begin{document}
 
\title{A Learning Based Scheme for Fair Timeliness in Sparse Gossip Networks}
 
\author{Purbesh Mitra \qquad Sennur Ulukus\\
        \normalsize Department of Electrical and Computer Engineering\\
        \normalsize University of Maryland, College Park, MD 20742\\
        \normalsize  \emph{pmitra@umd.edu} \qquad \emph{ulukus@umd.edu}}
\maketitle

\begin{abstract}
We consider a gossip network, consisting of $n$ nodes, which tracks the information at a source. The source updates its information with a Poisson arrival process and also sends updates to the nodes in the network. The nodes themselves can exchange information among themselves to become as timely as possible. However, the network structure is sparse and irregular, i.e., not every node is connected to every other node in the network, rather, the order of connectivity is low, and varies across different nodes. This asymmetry of the network implies that the nodes in the network do not perform equally in terms of timelines. Due to the gossiping nature of the network, some nodes are able to track the source very timely, whereas, some nodes fall behind versions quite often. In this work, we investigate how the rate-constrained source should distribute its update rate across the network to maintain fairness regarding timeliness, i.e., the overall worst case performance of the network can be minimized. Due to the continuous search space for optimum rate allocation, we formulate this problem as a continuum-armed bandit problem and employ Gaussian process based Bayesian optimization to meet a trade-off between exploration and exploitation sequentially.
\end{abstract}

\section{Introduction}\label{section: introduction}
Sparse gossip networks have emerged as a common network structure with the widespread use of internet of things (IoT) devices in different goal oriented applications~\cite{chettri2019comprehensive, swamy2020empirical}. Unlike dense networks, where each node in a network is connected to most of the other nodes, the IoT devices form heterogeneous structures in sparse networks and the order of connectivity varies greatly across different nodes. An example of such networks is the smart home environment, which combines controls of different devices across multiple platforms, such as television, air conditioner, door, refrigerator, lights, etc.~into a single controllable mobile application for a user. Another example is the sensor network inside a self-driving car, which contains multiple sensors across different positions in the car to effectively navigate in the real world. In these sparse networks, a common constraint for communication is bandwidth limitation. Since these networks are usually used in low power applications, it is crucial to have an optimized communication process in the network for effective bandwidth utilization to avoid \textit{information staleness} in the network. Staleness of information has been studied in different settings in the literature~\cite{kosta17AoIbook,yatesJSACsurvey,sun2022age}. In this work, we are interested in optimizing the source-to-network update rates of the connected devices to minimize an overall staleness metric. 

The first work to analyze network staleness in a gossip network is \cite{yates21gossip}, which has used version age of information as the metric for information freshness. In the system model, a source, with changing information versions over time, updates a network and the nodes in the network communicate with each other to track the source as timely as possible. The version age of an individual node at a certain time is defined as the difference between the version of information available at the node and the version of information at the source. \cite{yates21gossip} has derived a recursive formula for version age for fixed gossip rates using a stochastic hybrid system (SHS) analysis. 

\begin{figure}[t]
\centerline{\includegraphics[scale=0.7]{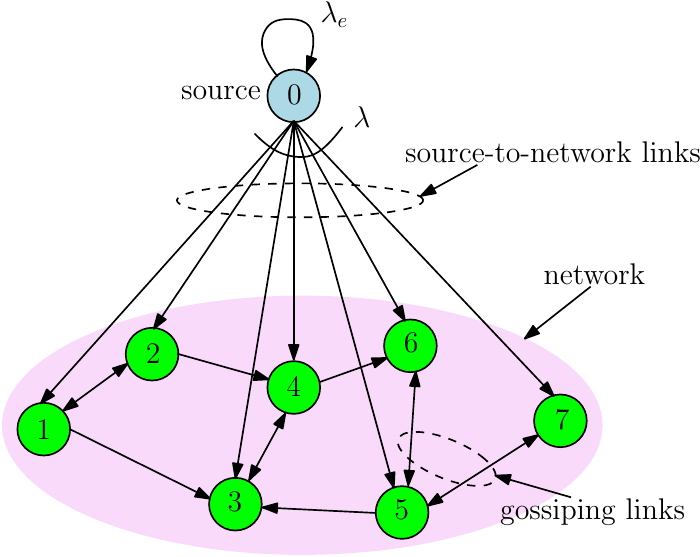}}
\caption{Sparse gossiping model with a $7$-node network tracking a source.}
\label{fig: system model}
\vspace*{-0.4cm}
\end{figure}

\begin{figure*}[t]
\centerline{\includegraphics[scale=0.7]{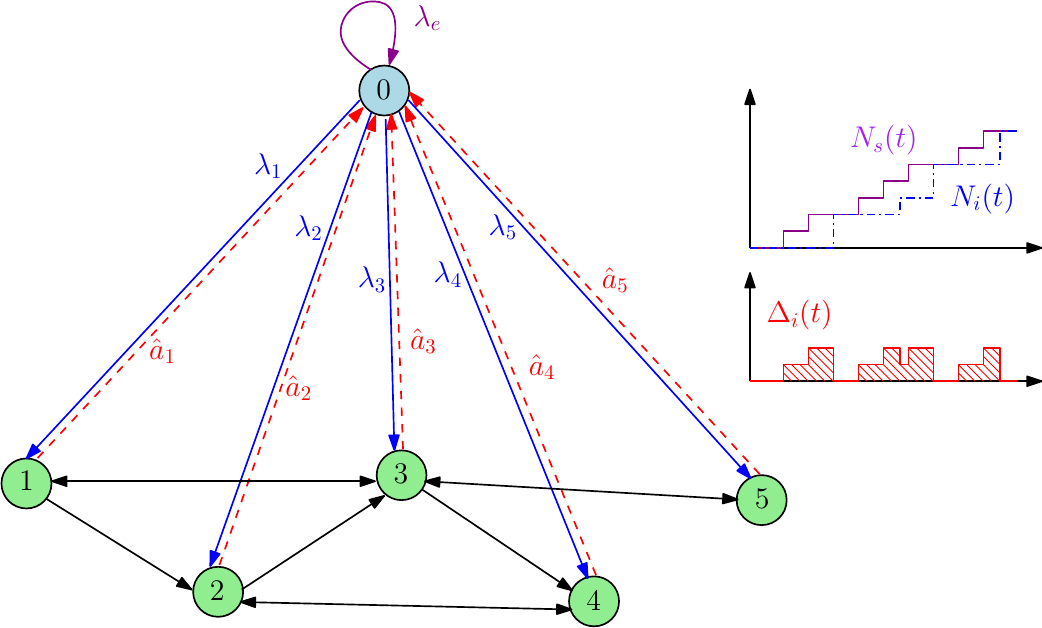}}
\caption{Learning based scheme for efficient allocation of update rates in a $5$-node sparse gossip network.}
\label{fig: learning update rate}
\end{figure*}

In the subsequent studies~\cite{buyukates22ClusterGossip,kaswan22slicingcoding,mitra_allerton22,mitra_infocom23}, the age performance has been improved using various techniques. \cite{buyukates22ClusterGossip} introduces a hierarchical network structure, in which nodes are clustered into different groups, where they can gossip among themselves. This improves the order of the age performance. In \cite{kaswan22slicingcoding}, the authors use network coding and file slicing for faster dissipation of generated data with high probability. The work in \cite{mitra_allerton22} introduces an age-aware gossiping technique that allows the fresh nodes to gossip with higher capacity and reduces the order of the age in a dense, fully connected network from $O(\log n)$ to $O(1)$, and \cite{mitra_infocom23} analytically shows the theoretical bound of age order reduction. Also see \cite{kaswan22jamming,kaswan22timestomp} for investigation of robustness/vulnerability of gossiping against adversarial actions such as jamming and timestomping.

In \cite{buyukates22ClusterGossip,kaswan22slicingcoding,mitra_allerton22,mitra_infocom23,kaswan22jamming,kaswan22timestomp}, the source-to-node structure is assumed to be symmetrical, and the gossip networks have either full $O(n)$ connectivity or very symmetric ring or hierarchical connectivity, which are easier to analyze. However, in real time scenarios, the network connectivity is not always guaranteed to have such ordered structure. Due to link failures, expensive bandwidth, physical separation of devices, the connectivity distributions vary greatly from one node to another. This poses a challenging problem in optimizing timeliness of the overall network as the analytical expression of age, introduced in \cite{yates21gossip}, depends on the whole network structure and gossip rates by a recursive formulation. Hence, the usual gradient-based optimization methods cannot be applied.

In this work, we introduce a learning based scheme to find a solution to the problem of sparse gossiping with fairness. The key idea of the proposed scheme is to assign different update rates from source to individual nodes and observe the effect on  the average age. Each node updates its neighbors according to their gossiping capacity. The nodes are assumed to be age-aware, i.e., they can keep track of their ages and calculate a time average. This time averaged age value is then sent back to the source for further sequential optimization. Then, the source applies a derivative free or black box optimization technique to fine-tune the update rates. The source, thus, acts as an orchestrator here alongside data generation and network updating. The source maintains fairness by minimizing the age of the worst performing node in the network to have an overall timeliness for the network. The learning algorithm will then dictate the new sets of gossip capacities, which will be used to evaluate the average ages in the next iteration. 

There are several derivative free optimization algorithms available in the literature~\cite{larson2019derivative, bubeck2011x}. An alternative way to formulate the problem is by considering the update rates as hyperparameters of some complex systems that can be optimized by some fine-tuning~\cite{shahriari2015taking}. In this work, we utilize Gaussian process based Bayesian optimization to minimize the overall worst case age of the network. In particular, we use the GP-UCB algorithm~\cite{srinivas2009gaussian} to formulate our problem as a continuum armed bandit problem and efficiently trade off between exploration and exploitation using the upper confidence bound. GP-UCB algorithm minimizes the expected regret over time, and under certain conditions guarantees sublinear regret asymptotically. This way, we use the performance guarantee of this algorithm to improve efficiency.

\section{System Model}\label{section: system model}
We consider a network, denoted as $\mathcal{N}$, that consists of $n$ nodes, which are tracking a source. The source updates its information with Poisson arrival rate of $\lambda_e$. The source updates nodes in the network with an overall Poisson arrival rate of $\lambda$. The nodes can gossip with their neighboring nodes and pass on information with the latest available version. The goal of the network is to track the source information as timely as possible. The timeliness of the $i$th node is characterized by its version age, which is the difference of version between the current information at the source and the information available at the node. Instantaneous version age can be written as
\begin{align}
    \Delta_i(t)=N_s(t)-N_i(t),
\end{align}
where $N_s$ and $N_i$ are the version processes at the source and node $i$, respectively. We denote the average version age as
\begin{align}
    a_i=\lim_{t\to\infty}\mathbb{E}[\Delta_i(t)].
\end{align}
From the analysis in \cite{yates21gossip}, for any set of nodes $\mathcal{S}\subset\mathcal{N}$, the average age is
\begin{align}\label{recursion eqn}
    a_{\mathcal{S}}\!=\!\!\lim_{t\to\infty}\!\mathbb{E}\!\!\left[\min_{i\in\mathcal{S}}\Delta_i(t)\!\right]\!\!=\!\!\frac{\lambda_e+\sum_{j\in N(\mathcal{S})}\lambda^{(j)}(\mathcal{S})a_{\mathcal{S}\cup\{j\}}}{\lambda^{(0)}(\mathcal{S})+\sum_{j\in N(\mathcal{S})}\lambda^{(j)}(\mathcal{S})},
\end{align}
where $N(\mathcal{S})$ denotes the set of neighboring nodes of $\mathcal{S}$, and $\lambda^{(0)}(\mathcal{S})$, $\lambda^{(j)}(\mathcal{S})$ are the arrival rates to $\mathcal{S}$ from the source and the $j$th node, respectively. Clearly, the average age of a node depends on the gossip rates among the nodes, the source-to-network update rates and the information generation rate $\lambda_e$ at the source. We denote the source to $i$th node update rate as $\lambda_i$. The individual average ages depend on all the $\lambda_i$'s due to the recursive function formulation in \eqref{recursion eqn} and the optimum choice of $\boldsymbol{\lambda}=\{\lambda_i\}_{i=1}^{n}$ would result in better timeliness performance for the network. Since we want to maintain an overall timeliness of the network subject to fairness, we minimize the average version age of the worst performing node $a(\boldsymbol{\lambda})=\max_{i\in\mathcal{N}}a_i(\boldsymbol{\lambda})$. Thus, the optimization problem is
\begin{align}\label{original optimization}
    \min_{\boldsymbol{\lambda}\in[0,\lambda]^n}&\quad a(\boldsymbol{\lambda})\notag\\
    \text{s.t.}&\quad\sum_{i=1}^{n}\lambda_i\leq\lambda.
\end{align}

\begin{algorithm}[t]
  \caption{Update rate learning algorithm}
  \label{algo: update rate learning}
  \begin{algorithmic}[1]
    \State Initialize $\boldsymbol{\lambda}_0=\left[\frac{\lambda}{n},\frac{\lambda}{n},\cdots,\frac{\lambda}{n}\right]$ at the source and start updating the network.
    \Procedure{SourceUpdate}{}
    \State Update the information at the source with rate $\lambda_e$.
    \State Send pilot signals to the network when updates occur.
    \EndProcedure
    \Procedure{NetworkUpdate}{}
    \State Update the nodes with rate $\boldsymbol{\lambda}$ from \textsc{SourceLearning}.
    \EndProcedure
    \For{$m\in[M]$}
        \Procedure{SourceLearning}{}
          \State Receive $\hat{a}_i$ from the \textsc{NodeFeedback}$(i)$.
          \State Wait until all the nodes have sent $\hat{a}_i$'s.
          \State Calculate $\mu_m(\boldsymbol{\lambda})$ and $\sigma_m(\boldsymbol{\lambda})$ using \eqref{mean regression} and \eqref{sigma regression}.
          \State Choose $\beta_m\sim O(\log (m^2))$.
          \State Get the acquisition function as
          \begin{align*}
              \phi_m(\boldsymbol{\lambda})=\mu_{m-1}(\boldsymbol{\lambda})+\sqrt{\beta_m}\sigma_{m-1}(\boldsymbol{\lambda}),
          \end{align*}
          \State Compute the new update rate distribution as
          \begin{align*}
              \boldsymbol{\lambda}_m=\text{arg}\max_{\boldsymbol{\lambda}\in \mathcal{D}}\phi_m(\boldsymbol{\lambda}).
          \end{align*}
        \EndProcedure
        \For{$i\in\mathcal{N}$}
            \Procedure{NodeFeedback}{node $i$}
                \State Receive updates from the source with rate $\lambda_i$.
                \State Gossip with its neighboring nodes.
                \State Keep track of the version age $\Delta_{i}(t)$.
                \State Calculate the time average age for a large time $T$ as
                \begin{align*}
                    \hat{a}_i=\frac{1}{T}\int_{T}\Delta_{i}(t)dt.
                \end{align*}
                \State Send $\hat{a}_i$ to the source.
            \EndProcedure
        \EndFor
    \EndFor
  \end{algorithmic}
\end{algorithm}

\section{Learning Based Scheme}\label{section: learning scheme}
In this section, we introduce a learning based scheme for fair allocation of the update rates. The recursive formulation in \eqref{recursion eqn} yields $n$ different formulations for $n$ nodes and taking a maxima over all of them is difficult to optimize for a general setting with gradient based methods, thus, we use an empirical strategy that does not rely on gradients and treat the age function as a noisy black box. We assume that the nodes are age sensing. This can be achieved by having a small pilot signal communication between the source and the network, whenever the source updates itself with new information. This way, the nodes become aware of the fact that there has been a change in the information version at the source. If the nodes are able to keep track of the number of such version updates at the source, they will be able to calculate their individual version ages. Consequently, the nodes calculate their time-averaged version ages and send feedback to the source.

The source decides the allocation of the update rate to the nodes sequentially, based on the average age performance. For this work, we consider the source as completely oblivious of the gossip network structure. It only receives the cost of its rate allocation from each individual node. The time average is taken over 
a long time window in each round. Hence, the source receives the estimates $\hat{a}_i=\frac{1}{T}\int_{T}\Delta_{i}(t)dt$, from the network, as shown in Fig.~\ref{fig: learning update rate}. Assuming ergodicity of the process, we have $\hat{a}_i\to a_i$ almost surely as $T\to\infty$, and therefore, for a large time window $T$, the estimate $\hat{a}_i\approx a_i$. Now, based on the sequential allocation-cost data, the source strategically explores the next allocation. Suppose, for a particular network, the optimum choice of rate allocation is denoted as $\boldsymbol{\lambda}^*$. We denote the $m$th choice of update rate allocation as $\boldsymbol{\lambda}_m$. With this choice, we can write the instantaneous regret as 
\begin{align}\label{inst regret}
    r_m=\hat{a}(\boldsymbol{\lambda}_m)-\hat{a}(\boldsymbol{\lambda}^*),
\end{align}
where $\hat{a}=\max_{i\in\mathcal{N}}\hat{a}_i$. The cumulative regret at the $M$th time step can be expressed as
\begin{align}\label{cumulative regret}
    R_M=\sum_{m=1}^{M}r_m.
\end{align}
To achieve asymptotic no-regret convergence to the optimum, the chosen sequence of rate allocations $\{\boldsymbol{\lambda}_m\}$ must yield
\begin{align}\label{sublinear regret}
    \lim_{M\to\infty}\frac{R_M}{M}=0\implies R_M= o(M),
\end{align}
i.e., a sublinear regret. Since the search space of this optimization problem is a continuous contour, this is a continuum-armed bandit problem. To solve this, we adapt the Bayesian optimization method with Gaussian process surrogate.

In this approach, we regress a Gaussian process (GP) with dimension $n$ to the age function as a surrogate for optimization. Then, we explore the search space with some acquisition function of our choice. In the Bayesian optimization literature, it is common to maximize some reward function instead of minimizing a cost function. Therefore, for simplicity, we formulate our original optimization problem in \eqref{original optimization} as a maximization of $f(\boldsymbol{\lambda})=-\hat{a}(\boldsymbol{\lambda})$. We regress the process $GP(\mu(\boldsymbol{\lambda}),k(\boldsymbol{\lambda},\boldsymbol{\lambda}'))$ to $f$, where $\mu(\boldsymbol{\lambda})$ is the mean of the process and $k(\boldsymbol{\lambda},\boldsymbol{\lambda}')$ is a regularized kernel. After performing the regression for $M$ time steps over the reward points $\boldsymbol{f}_M=[f_1, f_2, \cdots, f_M]^{T}$ corresponding to the sequence $\{\boldsymbol{\lambda}_m\}$, we obtain the mean and the variance as functions of the data as follows
\begin{align}
    \mu_M(\boldsymbol{\lambda})&=\boldsymbol{k}_M(\boldsymbol{\lambda})^{T}\boldsymbol{K}_M^{-1}\boldsymbol{f}_M,\label{mean regression}\\
    k_M(\boldsymbol{\lambda},\boldsymbol{\lambda}')&=k(\boldsymbol{\lambda},\boldsymbol{\lambda}')-\boldsymbol{k}_M(\boldsymbol{\lambda})^{T}\boldsymbol{K}_M^{-1}\boldsymbol{k}_M(\boldsymbol{\lambda}'),\label{kernel computation}\\
    \sigma^2_M(\boldsymbol{\lambda})&=k_M(\boldsymbol{\lambda},\boldsymbol{\lambda}),\label{sigma regression}
\end{align}
where $\boldsymbol{k}_M(\boldsymbol{\lambda})=[k(\boldsymbol{\lambda},\boldsymbol{\lambda}_1),k(\boldsymbol{\lambda},\boldsymbol{\lambda}_2),\cdots,k(\boldsymbol{\lambda},\boldsymbol{\lambda}_M)]^{T}$ and $\boldsymbol{K}_M$ is the kernel matrix
\begin{align}\label{kernel matrix}
    \boldsymbol{K}_M\!=\!\!\!
    \begin{bmatrix}
    k(\boldsymbol{\lambda}_1,\boldsymbol{\lambda}_1)&k(\boldsymbol{\lambda}_1,\boldsymbol{\lambda}_2)&\cdots&k(\boldsymbol{\lambda}_1,\boldsymbol{\lambda}_M)\\
    k(\boldsymbol{\lambda}_2,\boldsymbol{\lambda}_1)&k(\boldsymbol{\lambda}_2,\boldsymbol{\lambda}_2)&\cdots&k(\boldsymbol{\lambda}_2,\boldsymbol{\lambda}_M)\\
    \cdot&\cdot&\cdot&\cdot\\
    k(\boldsymbol{\lambda}_M,\boldsymbol{\lambda}_1)&k(\boldsymbol{\lambda}_M,\boldsymbol{\lambda}_2)&\cdots&k(\boldsymbol{\lambda}_M,\boldsymbol{\lambda}_M)
    \end{bmatrix}.
\end{align}

\begin{figure*}[t]
\subfigure[Update capacity $\lambda=1$ and $\lambda=2$.]{
\centering
\includegraphics[scale=0.55]{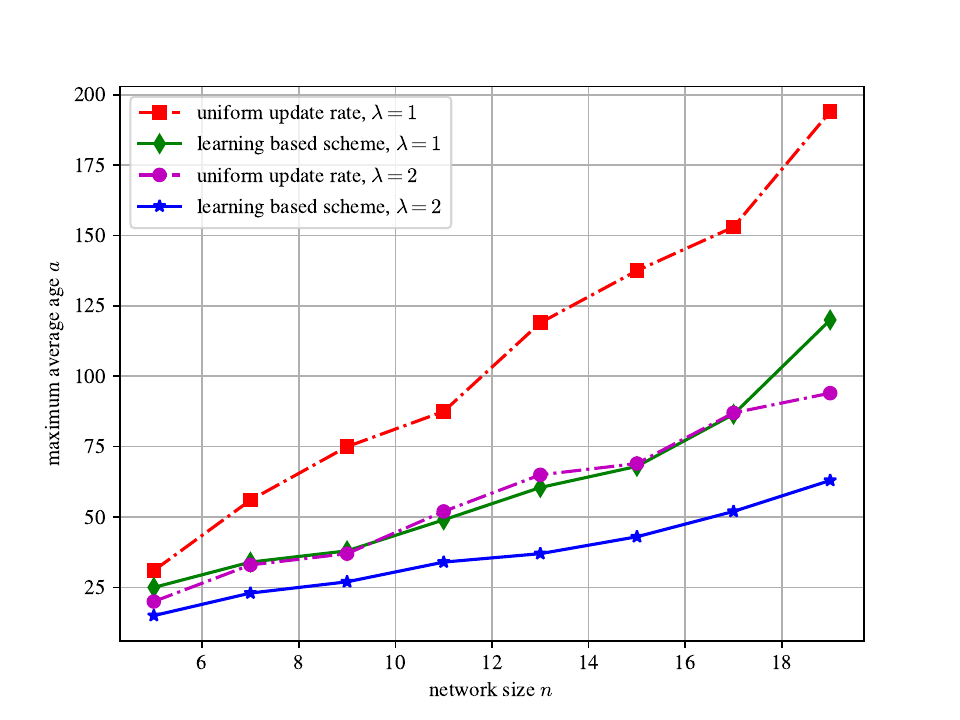}
\label{fig: plot_lambda}
}\hfill
\subfigure[Gossip capacity $B=\sqrt{n}$ and $B=2$.]{
\centering
\includegraphics[scale=0.55]{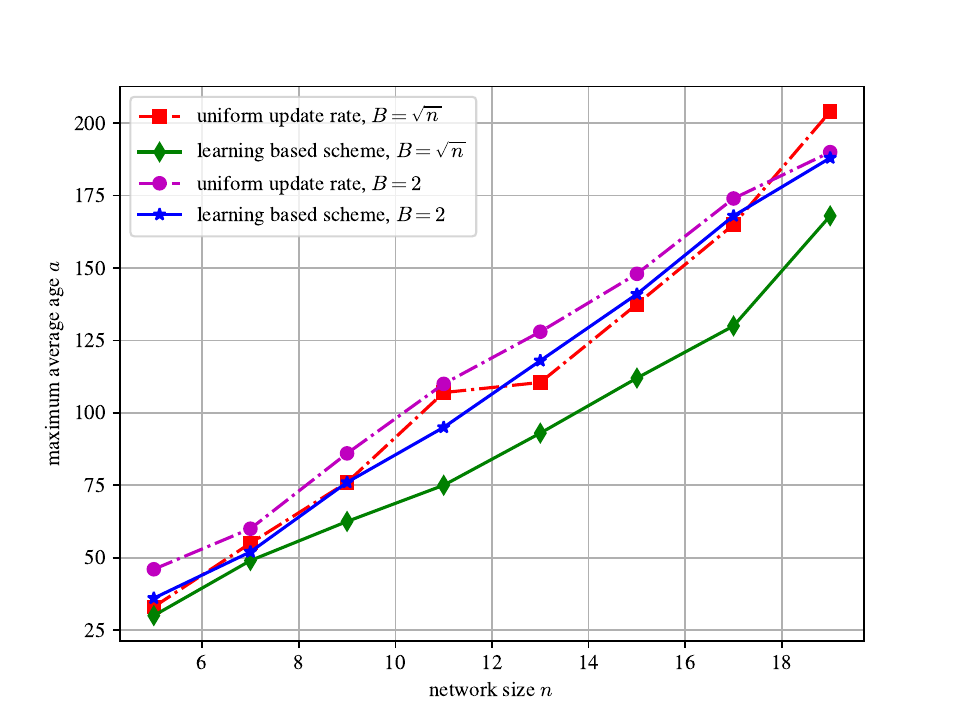}
\label{fig: plot_B}
}
\vspace*{-1mm}
\subfigure[$2$ and $3$ connections for each node.]{
\centering
\includegraphics[scale=0.55]{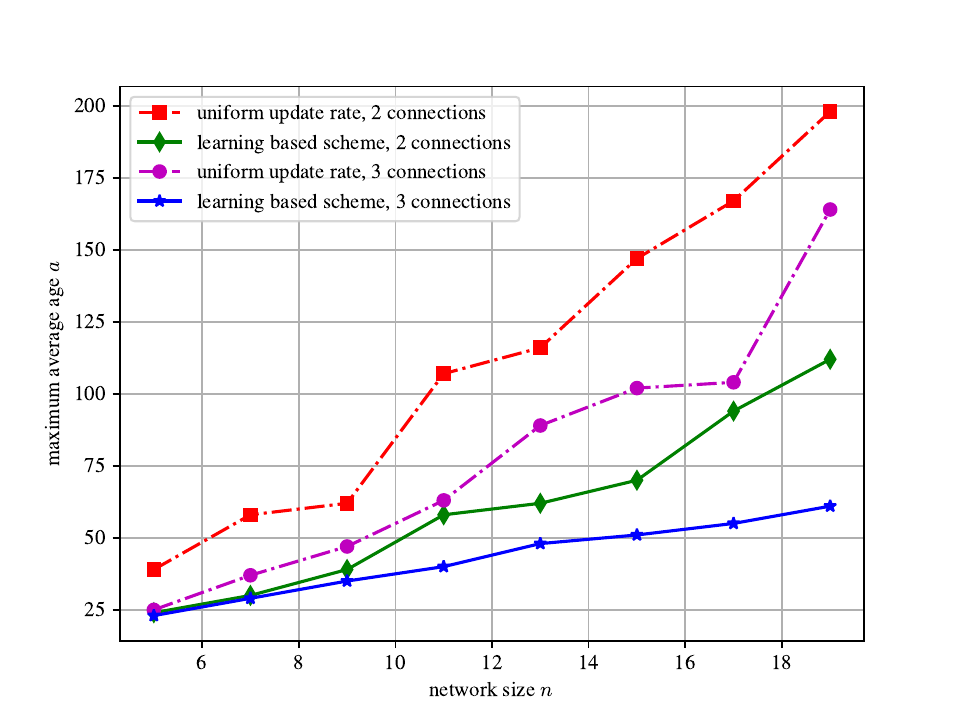}
\label{fig: plot_connections}
}\hfill
\subfigure[Exponential connectivity and random gossip connectivity.]{
\centering
\includegraphics[scale=0.55]{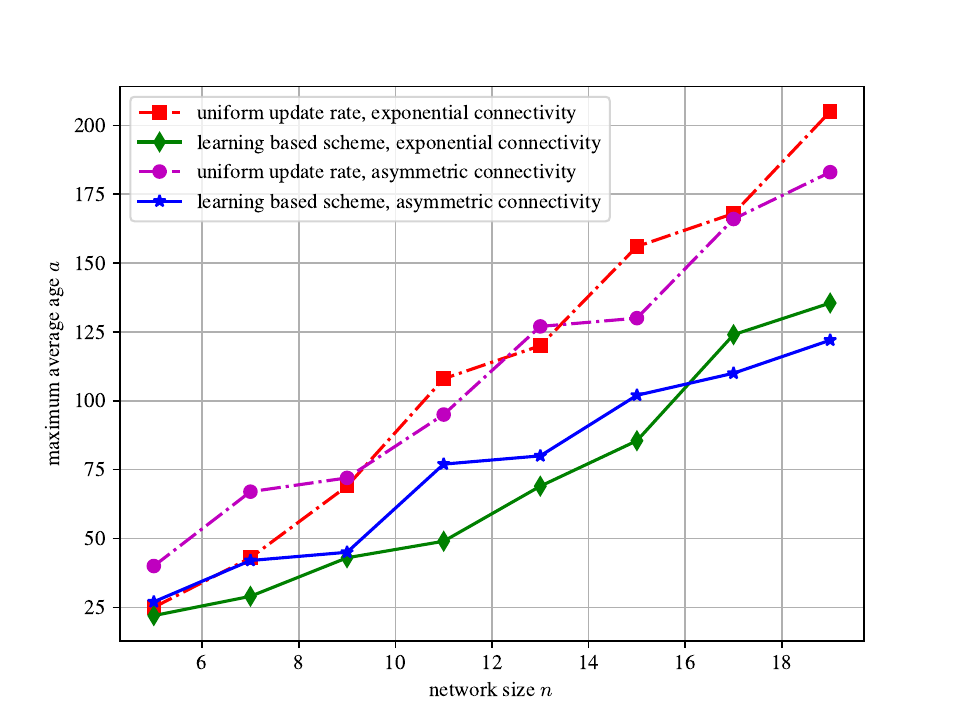}
\label{fig: plot_topology}
}
\caption{Learning based versus uniform update rate allocation for different constraints on update capacity, gossip capacity, number of connected neighbors, exponential connectivity topology and asymmetric gossip connections.}
\label{fig: comparison_update_gossip_connection}
\vspace*{-1mm}
\end{figure*}

\begin{figure*}[t]
\subfigure[A sample sparse graph with optimized age performance.]{
\centering
\includegraphics[scale=0.95]{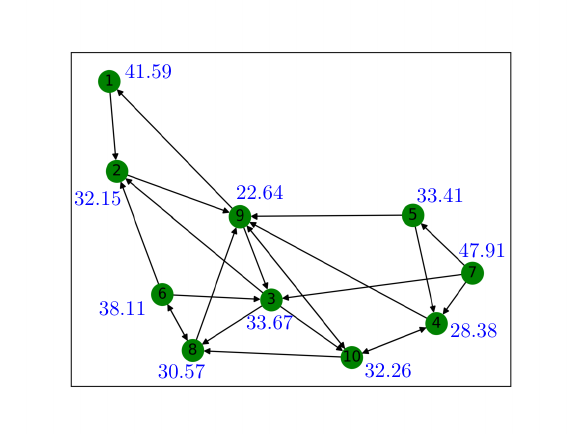}
\label{fig: sample_graph}
}\hspace{-5mm}
\subfigure[Histogram of average age of nodes.]{
\centering
\includegraphics[scale=0.46]{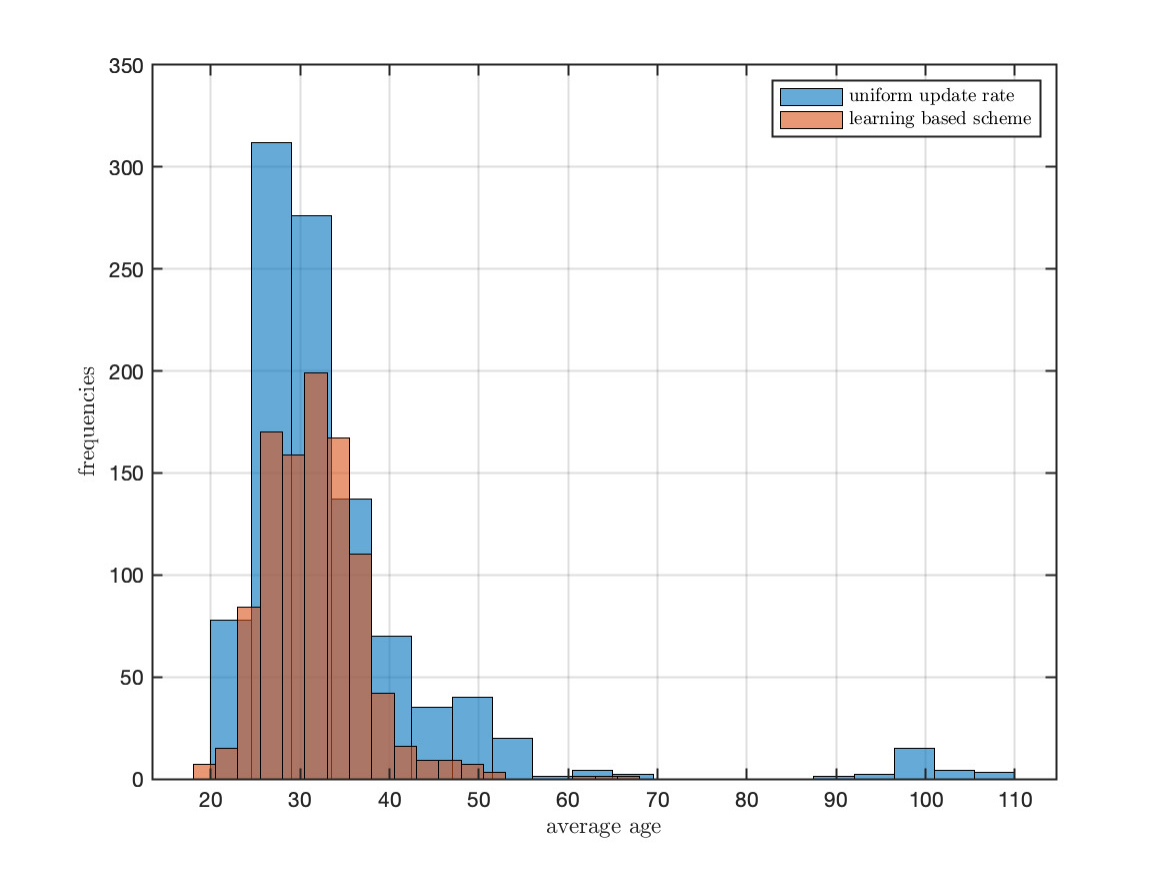}
\label{fig: histogram_age}
}
\vspace*{-1.1mm}
\subfigure[Average update rate versus outgoing connectivity.]{
\centering
\includegraphics[scale=0.55]{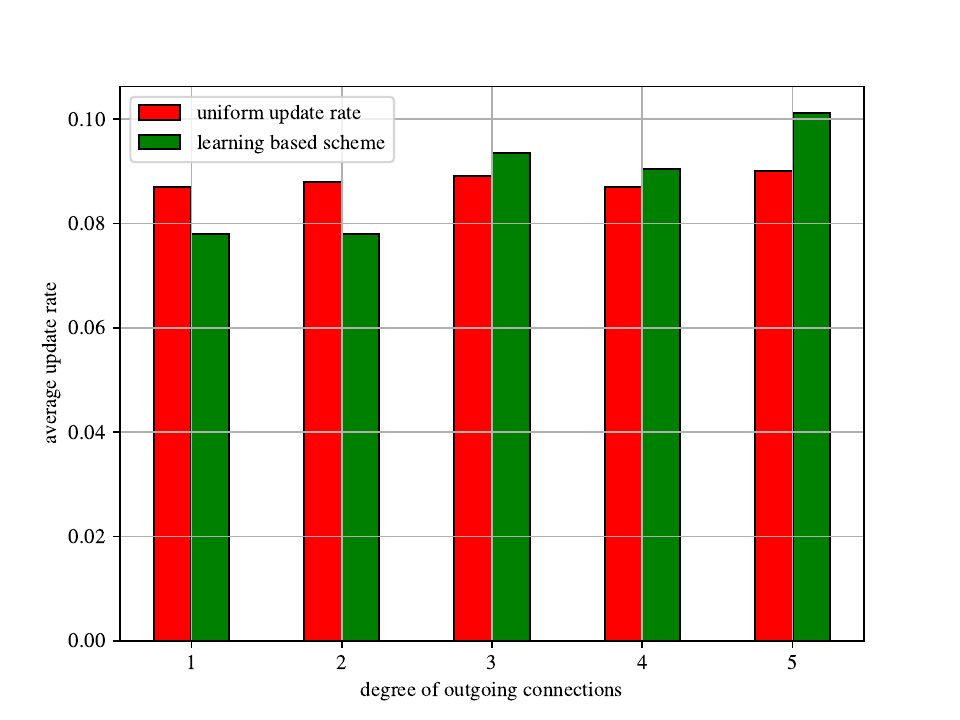}
\label{fig: out_deg}
}\hfill
\subfigure[Average update rate versus incoming connectivity.]{
\centering
\includegraphics[scale=0.55]{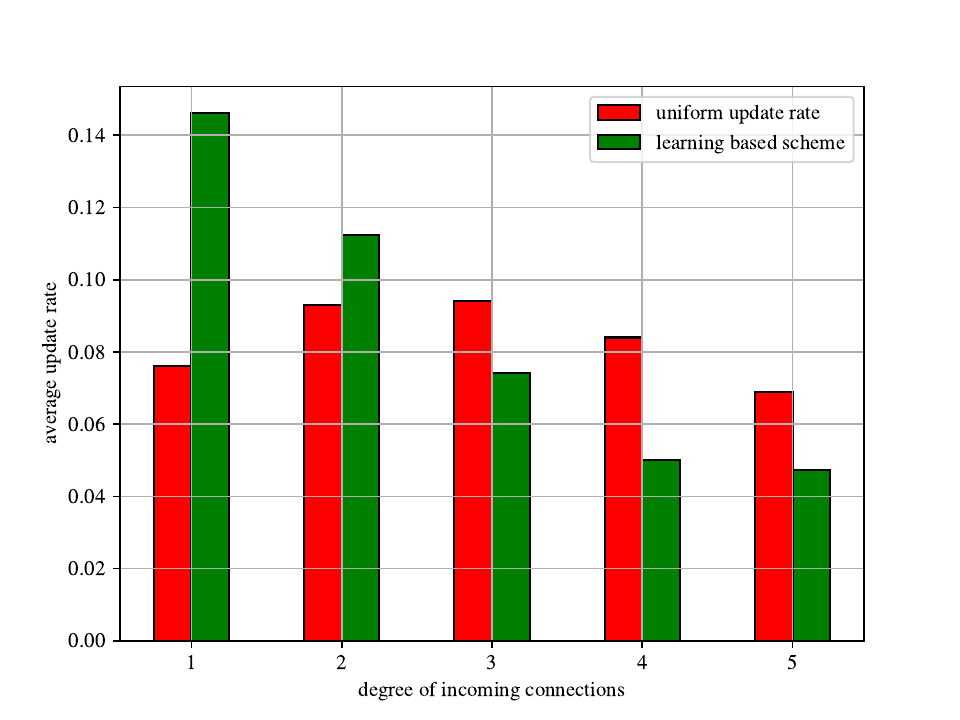}
\label{fig: in_deg}
}
\caption{Histogram of learning based versus uniform update rate allocation for different degrees of connectivity in gossip networks.}
\label{fig: comparison_hist}
\vspace*{-1.0mm}
\end{figure*}

Now, we choose an appropriate exploration strategy for generating the next step update rate allocation. One commonly used strategy is the upper confidence bound (GP-UCB) algorithm, which employs an optimization of a simple acquisition function over the search space. This auxiliary optimization is much simpler than the original optimization and can be performed with the existing gradient based numerical methods. The acquisition function for GP-UCB algorithm is
\begin{align}\label{acquisition function}
    \phi_m(\boldsymbol{\lambda})=\mu_{m-1}(\boldsymbol{\lambda})+\sqrt{\beta_m}\sigma_{m-1}(\boldsymbol{\lambda}),
\end{align}
where $\beta_m$ are weighting parameters, chosen appropriately. The term $\mu_{m-1}(\boldsymbol{\lambda})$ in \eqref{acquisition function} exploits information from the data points up to $m-1$ steps. Whereas, the term $\sigma_{m-1}(\boldsymbol{\lambda})$ pushes the algorithm for exploring different regions of the search space. At any given iteration, whether the algorithm will give more weightage to the exploration or the exploitation depends on the value of the $\beta_m$. This way, an exploration-exploitation trade-off is satisfied. By choosing
\begin{align}
    \boldsymbol{\lambda}_m=\text{arg}\max_{\boldsymbol{\lambda}\in \mathcal{D}}\phi_m(\boldsymbol{\lambda}),
\end{align}
where $\mathcal{D}=\{\boldsymbol{\lambda}\in[0,\lambda]^n:\mathbbm{1}^T\boldsymbol{\lambda}\leq \lambda\}$ is the search space, we obtain the update rate values to be used for worst case empirical age calculation in the next training cycle. Additionally, in \cite{srinivas2009gaussian}, the authors have analytically shown that for a low rank surrogate function $f$ in a convex compact search space, with the choice of $\beta_m\sim O(\log(m^2))$, the regret $R_M$ is guaranteed to be sublinear with arbitrary high probability, i.e., with high probability the algorithm converges to a global optimum. In our setting, $\mathcal{D}$ is a compact convex set, thus meeting the required criterion.

This process continues in the gossip network until the final update rates converge to an optimum. This learning based scheme is shown in Algorithm~\ref{algo: update rate learning}.

\section{Numerical Results}\label{section: numerical results}
In this section, we simulate different network topologies with selection of different parameters and compare the outcome of the learning based scheme with uniform rate allocation. We fix $\lambda_e=10$ and all the nodes are gossiping with total rate of $\frac{B}{n}$. The kernel $k$ used for simulation is Mat{\'e}rn kernel and the algorithm is run for $100$ iterations  for getting an optimal result.

First, we compare the effects of the update rate constraint $\lambda=1$ and $\lambda=2$. We fix $B=n$ for these simulations. The number of connections are chosen uniformly between $1$ and $3$ for the networks. From Fig.~\ref{fig: plot_lambda}, it is evident that with increasing update rate capacity, the age performance improves for both the uniform allocation and the optimized allocation. This is expected as having more source to network updates makes the overall network more timely. 

Next, we keep $\lambda=1$ and observe the effect of gossip capacity $B$ by varying it. We choose $B=\sqrt{n}$ and $B=2$ for the simulations. From Fig.~\ref{fig: plot_B}, we observe that for limited gossip capacity, the optimum update rate allocation is much closer to the uniform allocation. This is due to the fact that when the nodes in the network cannot gossip efficiently, the network becomes closer to a disconnected network, where each node only depends on updates from the source to be updated. Such a disconnected network has uniform update rate allocation as optimum due to its inherent symmetry. 

We also simulate the performance of the scheme for different sparse network configurations that are encountered in practical scenarios. We keep $\lambda=1$, $B=n$ and design networks for constant number of connections: $2$ and $3$ connections for each node. From Fig.~\ref{fig: plot_connections}, we observe that having more connections makes the optimum rate allocation perform better compared to uniform allocation. This can also be explained by the fact that increasing effective gossip makes the overall process more efficient. 

Finally, we simulate for exponential connectivity, i.e., number of connections of the $i$th node is $O(\gamma^i)$, and asymmetric gossip connectivity, where the gossip rates are not uniform across different nodes in Fig.~\ref{fig: plot_topology}. The learning based scheme manages to find an optimal solution in both of these cases.

More insights can be gained from Fig.~\ref{fig: comparison_hist} about the fairness of the update rate allocation. Fig.~\ref{fig: sample_graph} shows a sample directed graph, consisting of $10$ nodes. The minimum number of outgoing connections is $1$ and the maximum is $5$. The degree of each node is chosen in a uniformly random way between $1$ and $5$. Then, the learning based scheme is performed to optimize the overall worst case average age. From the figure, we observe that node $7$ has the worst case average age. This can be explained by the fact that node $7$ does not have any incoming connectivity and only receives updates from the source.

We plot a histogram of ages in randomly generated graphs with uniform update rates and the optimized update rates obtained by the learning based scheme in Fig.~\ref{fig: histogram_age}. We observe that for uniform update rate allocation, the histogram of average ages is more spread out in the plot, as compared to the learning based scheme, where the histogram is more concentrated, indicating that the difference between the smallest and the largest ages is smaller, which shows the fairness of the algorithm.

We also observe that due to the asymmetry of the connectivity in the directed graph, the final optimized age performance varies for different nodes. This motivates us to look for patterns for degree of outgoing and incoming connectivity versus the update rate allocations. We plot the average allocated update rate for different orders of outgoing and incoming degree distributions for randomly generated graphs. From Fig.~\ref{fig: out_deg}, we observe that for fair allocation, the learning based scheme assigns more update rate on average to the nodes with higher outgoing connectivity. This goes inline with intuition, as more outgoing connectivity of a node  generally relates to higher flow of information from that node, i.e., the node is significant for more outflow of fresh information to the network, and thus, assigned better update rate from the source. The opposite trend is noticed in Fig.~\ref{fig: in_deg}, as more average update rate is assigned to the nodes with fewer incoming connectivity. This also follows from the intuition, as low incoming connectivity is generally related to less incoming fresh information, thus, for a fair allocation the algorithm assigns more update rate to those nodes from the source.

\section{Conclusion}\label{section: conclusion}
We proposed a learning based scheme for source-to-network update rate allocation in a sparse gossip network. In a sparse network, due to heterogeneous connectivity, some nodes perform poorly compared to others in terms of timeliness. As a remedy, the source-to-node update rates can be chosen optimally such that the overall worst case performance is minimized. Due to the implicit interdependencies on the update rates to the nodes and gossiping, an analytical formulation for such problem is difficult to construct, hence, we proposed a learning based scheme for tuning the update rates. In particular, we adapted the continuum-armed bandit formulation of GP-UCB algorithm to design the scheme with some performance guarantee. In the scheme, the nodes are assumed to be age-sensing, and they send time-averaged version age after calculating for a large time as feedback to the source, which then uses the GP-UCB algorithm to find the optimum point by having a trade-off between exploration of the search space and exploitation of available information. 

\bibliographystyle{unsrt}
\bibliography{reference}

\end{document}